# A Novel Technique for Secret Message / Image Transmission through (2, 2) Visual Cryptographic Protocol (SMITVCP).

J.K. Mandal [1], S. Ghatak. [2]

[1] Department of Computer Science and Engineering, University of Kalyani, Kalyani, Nadia,
E-mail: jkm.cse@gmail.com

[2] A.K.C.S.I.T. University of Calcutta, Kolkata,
E-mail: subhankar.ghatak@gmail.com

**Abstract**

*In this paper a secret message/image transmission technique has been proposed through (2, 2) visual cryptographic share which is non-interpretable in general. A binary image is taken as cover image and authenticating message/image has been fabricated into it through a hash function where two bits in each pixel within four bits from LSB of the pixel is embedded and as a result it converts the binary image to gray scale one. (2, 2) visual cryptographic shares are generated from this converted gray scale image.*

*During decoding shares are combined to regenerate the authenticated image from where the secret message/image is obtained through the same hash function along with reduction of noise. Noise reduction is also done on regenerated authenticated image to regenerate original cover image at destination.*

I.   INTRODUCTION

In modern days image trafficking across the network, security is a big concern which can be achieved by steganography. Steganography is the art of secrete communication. The stganographic algorithms embed the secret information into different type of natural cover data like sound and images. The resulting altered data must be perceptually indistinguishable from its natural cover referred to as stego-data. The goal of steganography is to hide the message/image in the source image by some key techniques as the result observer has no knowledge of the existence of the message/image and it is unlike cryptography where the goal is to secure communications from an eavesdropper by making the data undetectable. As applications of steganography, the hidden data may be secrete message or secrete hologram whose mere presence within the host data set should be non-understandable. For image authentication and identification data hiding [1] in the image has become an important technique. The major task for research institute, scientist, and military people is ownership verification [8] and authentication. A technique for inserting information into an image for identification and authentication is known as image authentication. To protect digital image document from unauthorized access [9, 10] information security and image authentication has become very important. To hide a message/image inside an image without changing its visible properties [5] the source image may be altered. Chandramouli et al. [3] developed a useful method for making such alteration by masking, filtering and transformations of the least significant bit (LSB) on the source image. Dumitrescu et al. [4] construct an algorithm for detecting LSB steganography. From the recent works [7, 9] it is obvious that digital data can be effectively hidden in an image so as to satisfy the criteria that the degradation to host image is non-perceptible and it should be possible to recover the hidden message/image under a variety of attack. In this paper a new secrete message/image transmission scheme has been proposed where the secrete message/image has been embedded into the cover image through a hash function and the stego-data transmitted through (2,2) visual cryptographic protocol.

Visual cryptography is a new kind of security scheme that can be decoded directly by the human visual system without any special calculation for decryption. One of the nicest ones is the idea of secret sharing, originally suggested by Shamir [2]. The idea is to split a secret a (which can be a cryptographic key, but can also be any other piece of information) into n pieces (called shares), such that for s ≤ n (e.g., s = n). If an adversary has only s - 1 out of the n shares, then he has absolutely no information about the secret a. Given s shares it is possible to completely reconstruct the secret a. The proposed share generation technique is an extended visual cryptographic technique for gray scale image having sufficient band gap between black shaded and white shaded gray levels.

In this paper a message hyding technique has been proposed into the shares of visual cryptographic system. Section II deals with the proposed scheme SMITVCP. An example of SMITVCP is presented in section III. The correctness of decoding of SMITVCP is presented in section IV. Conclusions are given in section V.

II. THE SCHEME

The SMITVCP provides two-way security in secret message/image transmission. The cover image which is a binary image act as signature to authenticate the secret message/image and instead of normal post encryption transmission, the embedded image is again encrypted through (2, 2) VCS to act as one time pad to provide higher security level.

At transmission end, binary image is taken as cover image and authenticating message/image whose maximum size is (1/4) times of the cover image is taken (necessary condition). The proposed insertion and encoding algorithm embed each two consecutive bits of 8-bit binary representation of pixel/ASCII values of each character of authenticating image/message starting from MSB into two random positions of 8-bit binary representation of each pixel value of cover image. The positions are calculated using a hash function which is described in section A. The intermediate result is a gray scale embedded image on which (2, 2) VCS is applied to generate shares.

At the receiving end, shares are stacked in any order using the proposed decoding and extraction algorithm along with noise reduction technique. As an intermediate result the gray scale embedded image is regenerated and from this authenticated image the secret message/image is generated applying the same hash function. After extraction of authenticating message/image a second level of noise reduction is done on the extracted cover image to reconstruct the cover image which will be the exact binary image.

Section B deals with the insertion and encoding and that of section C describe decoding and extraction technique. The schematic diagram of the process is presented in figure 1.

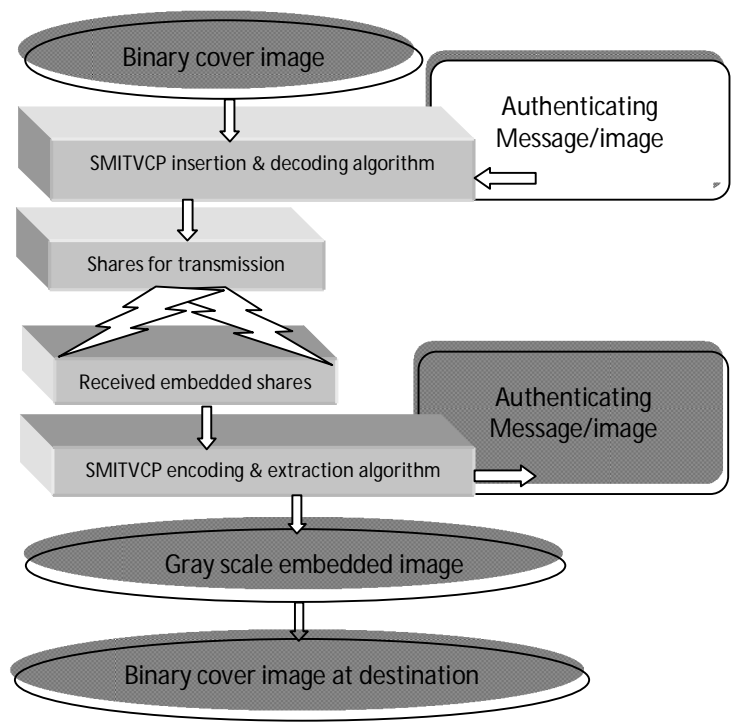

Figure 1: Schematic diagram of SMITVCP algorithm.

Department of Computer Science, The University of Burdwan, Burdwan, West Bengal, India

A. THE HASH FUNCTION

Hash function calculates the positions within the lower nibble of 8-bit binary representation of each pixel value of the cover image, where two consecutive bits starting from MSB of each pixel value/ASCII value of character of authenticating image/message are embedded.

Let P denotes the position of bit within 8-bit binary representation of the pixel value of cover image and p denotes the pixel number of cover image and s varies from 2 to 4. Considering the cover image in matrix form $I_{m \times n}$, where the size of the image is m × n and each pixel value is represented by $I_{ij}$, i be the number of row and j be the number of column. The mathematical expression for the function is given in equation 1:

$$\{P, P + 1\} = \{(\alpha = ((p \% ((i+1) + (j+1))) \% 4) \% s), (\alpha + 1)\}, P+1 = 0 \text{ if } P = 3 \quad \text{-------- (1)}$$

where, two bits from authenticating image/message are inserted into two positions P and P + 1 of each pixel $I_{ij}$ in $I_{ij}^{P}$ and $I_{ij}^{P+1}$ where superscripts indicating two bit positions in the lower nibble of the byte of $I_{ij}$.

B. THE INSERTION/ENCODING TECHNIQUE

Let $I_{m \times n}$ be a binary image taken as cover image, each pixel value of which is represented by $I_{ij}$ where i and j be the position of the pixel in terms of row and column respectively. Let $M_{x \times y}$ be the authenticating image/message (4 times smaller than the cover image in size).

Two consecutive bits $M_{00}^{7}$ and $M_{00}^{6}$ where superscripts indicating two bit positions calculated by the hash function in the lower nibble of byte of $I_{00}$.

As a result of embedding the binary cover image converted into a gray scale one. As next step of the scheme (2, 2) visual cryptographic protocol is applied on the embedded image to generate the shares. The proposed (2,2) visual cryptographic protocol generate two shares. In this process, the input is a gray scale image converted from binary image in the process of embedding. During insertion the pixel value 0 converted to the value say x where $0 \leq x \leq 12$ and the value 255 converted to the value say y where $243 \leq y \leq 255$. Considering the band (0, 12) as black pixel and (243,255) as white pixel. In the shares, the introduced noise in case of black pixel is 255 and that of white pixel is 0. The original pixel values of the embedded image are distributed to the shares using equation 2 and equation 3.

Let X be a pixel,

**If X be an even number,**
  Share1 = Floor(X/2) + 128,
  Share2 = Floor(X/2) + 128.
**If X be an odd number,**      ----- (2)
  Share1 = Floor(X/2) + 127,
  Share2 = {Floor(X/2) + 1} + 128.
where, $243 \leq X \leq 255$.

**If X be an even number,**
  Share1 = Floor(X/2),
  Share2 = Floor(X/2).
**If X be an odd number,**      ----- (3)
  Share1 = Floor(X/2),
  Share2 = {Floor(X/2) + 1}.
where, $243 \leq X \leq 255$.

A pixel can be shared by choosing one of the matrixes randomly obtained by permuting all columns. It has been considered that the bands (0-12) and (243-255) represent a black pixel and a white pixel respectively and denoted by '0' and '1' respectively. The corresponding matrices are as follows.

$C_0 = \left\{ \begin{pmatrix} 0 & 1 \\ 0 & 1 \end{pmatrix}, \begin{pmatrix} 1 & 0 \\ 1 & 0 \end{pmatrix} \right\}$ for white pixel and

$C_1 = \left\{ \begin{pmatrix} 0 & 1 \\ 1 & 0 \end{pmatrix}, \begin{pmatrix} 1 & 0 \\ 0 & 1 \end{pmatrix} \right\}$ for black pixel.

The pixel generation technique for the SMITVCP is given in the figure 2.

| Pixels | Share 1 | | Share 2 | | Probability of occurrence |
|---|---|---|---|---|---|
| Black pixel (0..12) | | | | | |
| 0 | 0 | 255 | 255 | 0 | |
| 1 | 0 | 255 | 255 | 1 | |
| …… | ……….. | ………. | ………. | | P = 1/13 |
| 12 | 6 | 255 | 255 | 6 | |
| 0 | 255 | 0 | 0 | 255 | |
| 1 | 255 | 0 | 1 | 255 | |
| …… | ……….. | ………. | ………. | | P = 1/13 |
| 12 | 255 | 6 | 6 | 255 | |
| White pixel (243..255) | | | | | |
| 243 | 248 | 0 | 249 | 0 | |
| 244 | 250 | 0 | 250 | 0 | |
| …… | ……….. | ………. | ………. | | P = 1/13 |
| 255 | 254 | 0 | 255 | 0 | |
| 243 | 0 | 248 | 0 | 249 | |
| 244 | 0 | 250 | 0 | 250 | |
| …… | ……….. | ………. | ………. | | P = 1/13 |
| 255 | 0 | 254 | 0 | 255 | |

Figure2: Share generation process of MITVCP

1. THE INSERTION/ENCODING ALGORITHM

**Input: An m × n cover image and authenticating message/image.**

**Output: Two shares, each of size m × 2n.**

Step 1. *Repeat for each character/pixel of authenticating message*

Step 2. *Generate 8-bit representation of ASCII value of character/pixel and put it into an array L(0-7).*

Step 3. *Generate 8-bit representation of each of 4 consecutive pixels of cover image and put it into an array $N_1(0-7), N_2(0-7), N_3(0-7), N_4(0-7)$.*

Step 4. *count ← 0; i ← 1;*

Step 5. *while (count ≤ 6) {*

Step 6. *Find the insertion positions P and P+1 for L(count) and L(count+1) respectively using equation 1.*

Step 7. *Replace $N_i(P)$ by L(count) and $N_i(P+1)$ by L(count+1);*

Step 8. *i ← i +1 and count ← count+2;*

Step 9. *al ← decimal value of $N_i(0-7)$;*

Step 10. *if(val ≥ 0 && val ≤12) then val is considered as black pixel.*

Step 11. *else if(val ≥ 243 && val ≤255) then val is considered as white pixel.*

Step 12. *Generate shares share1 and share2 using the proposed VCS protocol.*

Step 13. *} Stop.*

C. THE DECODING/EXTRACTION TECHNIQUE

Shares are received at destination end and authenticated image is regenerated by the following concept. Two consecutive pixel values are fetched from both shares say s1(1) & s1(2) and s2(1) and s2(2). Now if s1(1) is equal to s2(1) or s1(2) is equal to s2(2) then it represents a white pixel with noise 0. Otherwise it represents a black pixel with noise 255. The original pixel value is recovered from two pairs of pixel by the equation 4, equation 5 and equation 6 where noise is reduced automatically through the process:

If $Share1^1$ is equal to 0 and the value of $Share1^2$ is equal to $Share2^2$,
    value = $(Share1^2 + Share2^2) - 256$
If $Share1^1$ is equal to 0 and the value of $Share1^2$ is not equal to $Share2^2$,
    value = $(Share1^2 + Share2^2) - 254$
                                                -------- (4)
If $Share1^2$ is equal to 0 and the value of $Share1^1$ is equal to $Share2^1$,
    value = $(Share1^1 + Share2^1) - 256$
If $Share1^2$ is equal to 0 and the value of $Share1^1$ is not equal to $Share2^1$,

$$\text{value} = (Share1^1 + Share2^1) - 254 \quad \text{-------- (5)}$$

If $Share1^1$ is equal to 255,
$$\text{value} = (Share1^2 + Share2^1)$$
If $Share1^2$ is equal to 255,
$$\text{value} = (Share1^1 + Share2^2) \quad \text{-------- (6)}$$

Where superscripts denotes the pixel position within each group of two pixels in the corresponding share.

Form the group of four consecutive pixel values of regenerated authenticated image one character/pixel value of secret message/image is reconstructed by extracting two bits from each 8-bit representation of pixel value of authenticated image. The position of bits to be extracted is calculated by the same hash function. After extraction, a second level of noise reduction is done for the reconstruction of the original cover image from the authenticated image by the following proposed technique. Just replace the pixel value by 255 and 0 which is lies in the closed interval (243, 255) and (0, 12) respectively.

1. THE DECODING/EXTRACTION ALGORITHM

**Input:** Two shares, each of size m × 2n.
**Output:** The original cover image and an authenticating message/image.

*Step1. Do till the end of the share matrices.*

*Step2. Take two consecutive pixel values from each of the share.*

*Step3. Evaluate whether these represents one black pixel or white pixel using the above mentioned concept.*

*Step4. If it is a white pixel then discard the 0 values and calculate the original pixel value from other pixel values by the equation 4 and equation 5.*

*Step5. Else discard the 255 values and calculate the original pixel value from other pixel values by the equation 6.*

*Step6. Repeat for each group of four pixel values of the reconstructed authenticated image.*

*Step7. Calculate 8-bit representation of each of 4 consecutive pixels of authenticated image and put it into an array $N_1(0-7)$, $N_2(0-7)$, $N_3(0-7)$, $N_4(0-7)$.*

*Step8. count ← 0; i ← 1;*

*Step9. While (count ≤ 6) {*

*Step10. Find the extraction positions P and P+1 for L(count) and L(count+1) respectively using equation 1 where L(0-7) representing each character/pixel of authenticating message/image .*

*Step11. Replace L(count) by $N_i(P)$ and L(count+1) by $N_i(P+1)$ ;*

*Step12. If( 243 ≤ decimal value of $N_i(0-7)$ ≤ 255) then store the value 255 in data storage 1.*

*Step13. Else if( 0 ≤ decimal value of $N_i(0-7)$ ≤ 12) then store the value 0 in data storage 1.*

*Step14. i ← i +1 and count ← count+2;}*

*Step15. Store the corresponding character/pixel value of L(0-7) in data storage 2.*

*Step16. Stop.*

III. THE DECODING CORRECTNESS

In visual cryptography decoding is done directly by human visual system. In SMITVCP decoding is done through some cryptographic computation. But the secret authenticated image can also revel by human visual system if shares are printed on different transparent sheets and staked them with proper alignment. If a transparency printed by one gray level is superimposed with another gray level then the color will seen as resultant color. The gray level is proportional to the Humming weight [12] H(S) and interpreted as black if H(S) ≥ 2 and as white if H(S) ≤ 1. The resultant colors are generated by stacking the shares can be represented by the logical OR operation. The Humming weight of the ORed pixel can be evaluated by H (Share1 OR Share2).

In SMITVCP the cover image is a binary image and after insertion operation to converts to gray scale image on which proposed (2, 2) VCS is applied, with the help of huge band gap (13-244) available between black shaded pixel and white shaded pixel it can be easily identify the type (B/W) of the pixel. Figure 4 depicts the Histogram of cover image and authenticated image respectively.

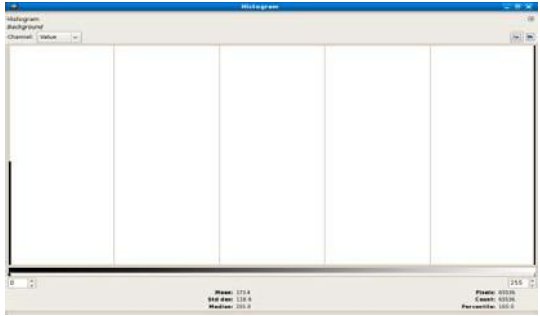

(a)

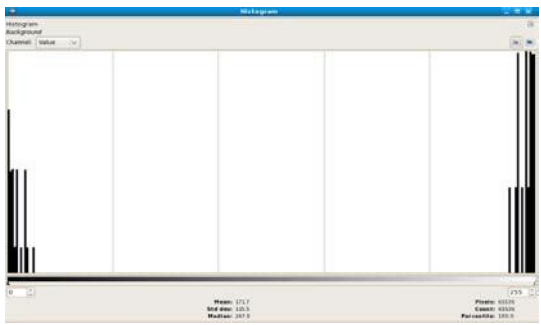

(b)

Figure 4: Histogram of (a) Cover image (Binary), (b) Authenticated image (Grayscale).

From the shares the original value can be evaluated by the following expression ((Share1(1) AND Share2(1)) + (Share1(2) AND Share2(2))) for black pixels (0-12) and for white pixel (243-255) if value of Share1(1) is equal to 0 then if value of Share1(2) is equal to value of Share2(2) expression is ((Share1(2) + Share2(2)) – 256) else ((Share1(2) + Share2(2)) – 254) otherwise if value of Share1(1) is equal to value of Share2(1) expression is ((Share1(1) + Share2(1)) – 256) else ((Share1(1) + Share2(1)) – 254). Using the above expressions the evaluated pixel value is positioned wise same as the original pixel value which conforms that the regenerated image and the authenticated image are same.

## IV. THE EXPERIMENTAL RESULTS AND COMPARISONS

The image given in figure 5 and 6 is used for simulation of the proposed SMITVCP and the resultant images are given in figure 7, 8, 9. If we stack figure 8 and 9 we get the same figure presented in figure 7.

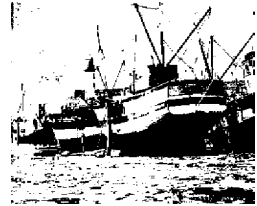 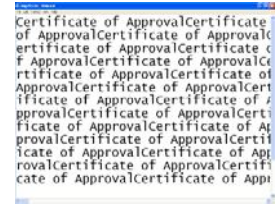

Figure 5: Cover Image     Figure 6: Secret message

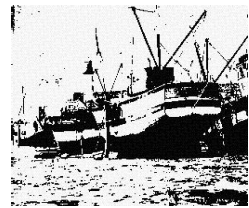 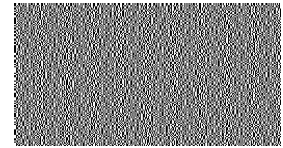

Figure 7: Embedded image.     Figure 8: Share 1 of embedded image (fig. 7)

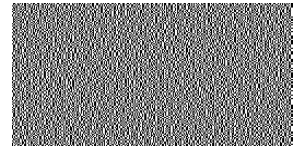

Figure 8: Share 2 of embedded image (fig. 7)

The comparison of the proposed (2, 2) VCS in SMITVCP with other existing VCS in terms of probability of occurrence is drown in figure 9.

| Sl. No. | Algorithms | Probability of occurrence |
|---|---|---|
| 1. | Visual Cryptography by Moni Naor and Adi Shamir [2]. | 1/2 |
| 2. | A Neural Network Approach for Visual Cryptography by Tai-Wen Yue and Suchen Chiang [6]. | 1/2 |
| 3. | A Novel Visual Cryptography scheme by Debasish Jena, Sanjay Kumar Jena [11]. | 1/2 |
| 4. | SMITVCP (The proposed technique) | 1/13 |

## V. CONCLUSION

In this paper a new cryptographic technique has been proposed where in addition to generating and transmitting shares, secret message transmission has also been introduced. Proposal may enhance the level of security further and alternative secrete way of message transmission may be ensured.

ACKNOWLEDGMENT

REFERENCE

1. Amin, P., Lue, N. and Subbalakshmi, K., "Statistically secure digital image data hiding", in *IEEE Multimedia Signal Processing MMSP05*, China, Oct. 2005, pp. 1-4.

2. Naor, M., Shamir, A.: Visual Cryptography. In De Santis, A. (ed.) EUROCRYPT 1994. LNCS, vol. 950, pp. 1-12. Springer, Heidelberg (1995).

3. Chandramouli, R. and Memon, N., "Analysis of LSB based image steganography techniques", in *Proc. of ICIP*, Thissaloniki, Greece, 2001, pp. 1019-1022.

4. Dumitrescu, S., Xiaolin, W. and Wang, Z., "Detection of LSB steganography via sample pair analysis", In: *LNCS*, Springer-Verlog, New York, 2003, Vol. 2578, pp: 355-372.

5. Moulin, P. and O'Sullivan, J.A., "Information Theoretic analysis of information hiding", in *IEEE Trans. on Info. Theory*, March 2003, Vol. 49, No. 3, pp. 563-593.

6. Yue, T.W., Chiang, S.: A Neural Network Approach for Visual Cryptography. In: IEEE-INNS-ENNS International Joint Conference on Neural Networks, vol. 5, pp. 494-499 (2000)

7. Nameer, N. EL-Emam, " Hiding a large amount of data with high security using steganography algorithm", *Journal of Computer Science ISSN 1549-3636,* Vol. 3, No. 4, 2007, pp: 223-232.

8. Paven, S., Gangadharpalli, S. and Sridhar, V., "Multivariate entropy detector based hybrid image registration algorithm", in *IEEE Int. Conf. on Acoustic, Speech and Signal Processing,* Philadelphia, Pennsylvania, USA, March 2005, pp:18-23.

9. Pang, H.H., Tan, K.L. and Zhou, X., "Steganographic schemes for file system and b-tree", in *IEEE Trans. On Knowledge and Data Engineering,* Singapore, June 2004, Vol.16. pp: 701-713

10. Rechberger, C., Rijman V. and Sklavos N., "The NIST cryptographic Workshop on Hash Functions", in *IEEE Security & Privacy,* Austria, Jan/Feb 2006, Vol. 4, pp. 54-56.

11. Jena, D., Jena, S.K.: A Novel Visual Cryptography Scheme. In: The 2009 International Conference on Advance Computer Control, pp. 207-211 (2009).

12. Gravano, S.: Introduction to Error Control Codes. Oxford University Press, USA (2001).